\documentclass[useAMS,usenatbib]{mnras}

\usepackage[utf8]{inputenc}
\usepackage{amssymb,amsmath}
\usepackage{graphicx}
\usepackage[dvipsnames]{xcolor}
\hypersetup{colorlinks=true,allcolors=violet}
\usepackage{todonotes}

\newcommand{\be}{\begin{equation}}
\newcommand{\ee}{\end{equation}}
\def\bear#1\ear{\begin{align}#1\end{align}}
\newcommand{\nline}{\notag \\}
\newcommand{\f}{\frac}
\newcommand{\de}{\mathrm{d}}

\newcommand{\e}{\mathrm{e}}

\renewcommand{\mathbf}[1]{\mbox{\boldmath $#1$}}

\newcommand{\Msun}{\mathrm{M}_{\odot}}

\newcommand{\fig}[1]{Figure \ref{#1}}

\title[Reionization constraints from CMB]{CMB constraints on a physical model of reionization}
\author[Choudhury, Mukherjee \& Paul]{
Tirthankar Roy Choudhury$^1$\thanks{tirth@ncra.tifr.res.in},
Suvodip Mukherjee$^{2}$\thanks{s.mukherjee@uva.nl, mukherje@iap.fr},
Sourabh Paul$^{3}$\thanks{sourabh.paul@gmail.com}
\\
$^{1}$ National Centre for Radio Astrophysics, Tata Institute of Fundamental Research, Pune 411007, India\\
$^{2}$ Gravitation Astroparticle Physics Amsterdam (GRAPPA),
Anton Pannoekoek Institute for Astronomy and Institute for High-Energy Physics,\\
University of Amsterdam, Science Park 904, 1090 GL Amsterdam, The Netherlands\\
$^{3}$ Department of Physics and Astronomy, University of the Western Cape, Bellville, Cape Town, South Africa\\
}
\begin{document} 
 
\date{} 

\maketitle

\begin{abstract}
We study constraints on allowed reionization histories by comparing predictions of a physical semi-numerical model with secondary temperature and polarization anisotropies of the cosmic microwave background (CMB).
Our model has four free parameters characterizing the evolution of ionizing efficiency $\zeta$ and the minimum mass $M_{\mathrm{min}}$ of haloes that can produce ionizing radiation.
Comparing the model predictions with the presently available data of the optical depth $\tau$ and kinematic Sunyaev-Zeldovich signal, we find that we can already rule out a significant region of the parameter space.
We limit the duration of reionization $\Delta z= 1.30^{+0.19}_{-0.60}$ ($\Delta z < 2.9$ at $99\%$ C.L.), one of the tightest constraints on the parameter. 
The constraints mildly favour $M_{\mathrm{min}} \gtrsim 10^9 \Msun$ (at $68\%$ C.L.) at $z \sim 8$, thus indicating the presence of reionization feedback. 
Our analysis provides an upper bound on the secondary $B$-mode amplitude $D_{l=200}^{BB} < 18$ nK$^2$ at $99\%$ C.L.
We also study how the constraints can be further tightened with upcoming space and ground-based CMB missions.
Our study, \emph{which relies solely on CMB data}, has implications not only for upcoming CMB surveys for detecting primordial gravitational waves but also redshifted 21~cm studies.
\end{abstract}


%#################################################################
\section{Introduction} 
%#################################################################
Reionization of cosmic neutral hydrogen (HI) by the first stars provides a natural method to study the high-redshift universe. The cosmic microwave background (CMB) provides an exquisite window to explore the reionization history of HI using the secondary temperature and polarization anisotropies induced during its propagation from the surface of the last scattering \citep{Sugiyama1993}. Conventional methods of constraining reionization using the CMB implement rather simple parametrizations of the reionization history, e.g., using the mean redshift and the duration of reionization \citep[see, e.g.,][]{Battaglia2013,PlanckCollaboration2018,Reichardt2020}. As per our current understanding, the ionization field during reionization is expected to be ``patchy'', characterized by overlapping bubbles around the galaxies. The models, based on these simple parametrizations, ignore the dependence of the CMB observables on the patchiness which are known to play significant role \citep{Mukherjee2019,Roy2020,Paul2020}.

Our main aim is to build on the existing analyses and use a physically motivated model to put constraints on reionization history by comparing with \emph{only CMB observables}. The advantage of using a physical model is that it allows connecting the resulting constraints with the physics of the high-redshift Universe. Note that such models have been widely used to constrain reionization by comparing with CMB and other observations, e.g., Ly$\alpha$ absorption at $z \sim 6$ \citep[for recent results, see e.g.,][]{Mitra2018,Qin2020}. However, the recent improvements in CMB data (and more expected in the near future), it becomes useful to check how effective the CMB experiments are in studying reionization. It is with this aim that we restrict our analysis to only CMB observables, although our formalism is well adapted to be applied to other observations too.

%\vspace{-1cm}
Our analysis is divided into two parts: In the first and main part, we constrain the reionization history using presently available observations, namely, the optical depth $\tau$ measurements from Planck \citep{PlanckCollaboration2018} and the kinematic Sunayeav Zeldovich (kSZ) signal from the South Pole Telescope \citep[SPT,][]{Reichardt2020}. The aim here is to understand if there is a class of models that can already be ruled out. In the second part, we extend our analysis to make forecasts for ongoing and upcoming CMB probes, e.g., the upcoming space-based mission LiteBIRD \citep{Suzuki2018} and the ground-based CMB experiments Atacama Cosmology Telescope \citep[ACT,][]{Henderson2016}, Simons Observatory \citep{Ade2019} and CMB-S4 \citep{Abazajian2019}.

%\vspace{-0.5cm}
\section{Simulations and CMB data}
\label{sec:method}
The ionization maps needed for this work are generated using the photon-conserving semi-numerical scheme \texttt{SCRIPT} (\textbf{S}emi-numerical \textbf{C}ode for \textbf{R}e\textbf{I}onization with \textbf{P}ho\textbf{T}on-conservation), for details, see \citet{Choudhury2018,Choudhury2020}. We use GADGET-2 \citep{Springel2005} to generate the large-scale matter density and velocity fields in a box of length $512 h^{-1}$~cMpc with $256^3$ particles.\footnote{The cosmological parameters used in this work are $\Omega_m = 0.308, \Omega_{\Lambda} = 1- \Omega_m, \Omega_b = 0.0482, h = 0.678, n_s = 0.961, \sigma_8 = 0.829$ \citep{PlanckCollaboration2014}.} We simulate the distribution of haloes using a sub-grid prescription based on the ellipsoidal collapse \citep{Sheth2002}. Our method produces halo mass functions consistent with the $N$-body simulations of \citet{Jenkins2001}. At a given redshift $z$, \texttt{SCRIPT} takes two input parameters, namely, the ionizing efficiency $\zeta$ of star-forming haloes and the minimum mass $M_{\mathrm{min}}$ of haloes which can produce ionizing photons, and outputs the ionized hydrogen fraction $x_{\mathrm{HII}}(\mathbf{x}, z)$ for each grid cell in the simulation volume. Of interest to us is the free electron fraction $x_{e}(\mathbf{x}, z) = \chi_{\mathrm{He}}(z) ~ x_{\mathrm{HII}}(\mathbf{x}, z),$ where $\chi_{\mathrm{He}}$ accounts for the excess electron correction factor due to ionized Helium.\footnote{We assume $\chi_{\mathrm{He}} = 1.08$ for $z > 3$ (singly ionized Helium) and $\chi_{\mathrm{He}} = 1.16$ for $z \leq 3$ (double ionized Helium).} Both $\zeta$ and $M_{\mathrm{min}}$ are determined the galaxy formation physics at high redshifts and their evolution is not straightforward to model. In the absence of any insights at high redshifts, they are often taken to be independent of redshift \citep[see, e.g.,][]{Mesinger2012}. In this work, we  assume both $\zeta$ and $M_{\mathrm{min}}$ to have power-law dependencies on $z$
\be
\zeta(z) = \zeta_0 \left(\f{1 + z}{9}\right)^{\alpha_{\zeta}}, M_{\mathrm{min}}(z) = M_{\mathrm{min},0} \left(\f{1 + z}{9}\right)^{\alpha_{M}},
\ee
where $\zeta_0$ and $M_{\mathrm{min},0}$ are the values at $z = 8$. To keep the number of parameters under control, we ignore any mass-dependence of $\zeta$. Hence, our reionization model is fully described by four free parameters.
 
The Thomson scattering of the CMB quadrupole by the free electrons available during the epoch of reionization (EoR) leads to secondary $E$-mode polarization signal at low-$l$, which can be quantified in terms of the optical depth $\tau \equiv \tau(z_{\mathrm{LSS}})$ to the last scattering redshift $z_{\mathrm{LSS}}$, where 
\be
\tau(z) = \sigma_T \bar{n}_{H}c \int_0^{z} \f{\de z'}{H(z')} ~ (1 + z')^2 ~ \chi_{\mathrm{He}}(z') ~ Q_{\mathrm{HII}}(z').
\label{eq:tau_z}
\ee
Above, $\bar{n}_H$ is the mean comoving number density of hydrogen, $\sigma_T$ is the Thomson cross section and $Q_{\mathrm{HII}}(z)$ is the mass-averaged ionized fraction obtained from \texttt{SCRIPT}.

The kSZ signal during EoR arises from the bulk motion of the ionized bubbles with respect to the CMB and the relevant quantity is the dimensionless momentum field 
$\mathbf{q}(\mathbf{x}, z) \equiv x_{e}(\mathbf{x}, z) \Delta(\mathbf{x}, z) \mathbf{v}(\mathbf{x}, z) / c$. Under Limber's approximation, the kSZ angular power spectrum can be estimated using \citep{Ma2002, Mesinger2012, Park2013, Alvarez2016}
\bear
C_l^{\mathrm{kSZ,patchy}} &= \left(\sigma_T \bar{n}_{H} T_0\right)^2 \int_0^{z_{\mathrm{LSS}}} \f{c~\de z}{H(z)}~\f{(1+z)^4}{\chi^2(z)} \times
\nline
& \times \e^{-2 \tau(z)}~\f{P_{q_\perp}(k = l/\chi(z), z)}{2},
\label{kszeq}
\ear
where $T_0 = 2.725$ K is the present CMB temperature, $\chi(z)$ is the comoving distance to $z$, and $P_{q_\perp}(k, z)$ is the power spectrum of the transverse component of the Fourier transform $\mathbf{q}(\mathbf{k}, z)$ of the momentum field defined as $\mathbf{q_\perp}(\mathbf{k}, z) = \mathbf{q}(\mathbf{k}, z) - \left(\mathbf{q}(\mathbf{k}, z) \cdot \mathbf{k}\right) \mathbf{k} / k^2$. 

The observed kSZ is an integrated effect that gets contribution from both during and post reionization epochs. During post-reionization, the signal is sourced by the Ostriker-Vishniac (OV) effect \citep{Ostriker1986, Ma2002}, which requires modelling of the non-linear density and velocity fields \citep{Shaw2012}. While comparing the models with data, we add the OV contribution to that from patchy reionization using the scaling laws given in \citet{Shaw2012}. Hence the total kSZ power spectrum can be computed as $C_l^{\mathrm{kSZ,tot}} = C_l^{\mathrm{kSZ, OV}} + C_l^{\mathrm{kSZ, patchy}}$. 

%\vspace{-0.5cm}
\subsection{Data sets and likelihood}

The best constraints on $\tau$ at present comes from low-$l$ $E$-mode polarization from Planck, given by $\tau^{\mathrm{obs}}= 0.054$ with error $\sigma_{\tau}^{\mathrm{obs}} = 0.007$ \citep{PlanckCollaboration2018}. For the kSZ power spectrum, we use the first statistically significant detection reported by the SPT as $D^{\mathrm{kSZ, obs}}_{l=3000} \equiv l(l+1)C^{\mathrm{kSZ, obs}}_l/2\pi= 3 \mu \mathrm{K}^2$ with a standard deviation $\sigma^{\mathrm{kSZ}}_{l=3000} = 1\mu \mathrm{K}^2$ \citep{Reichardt2020}.

\begin{table}
\begin{tabular}{|c|c|c|c|c|}
\hline
Mission & Frequency & $\Delta T$ & Beam & $f_{\mathrm{sky}}$\\ 
        & (GHz) & ($\mu \text{K}$-arcmin) & (arcmin)& \\ 
\hline
Adv-ACTPol& 150  &7 & $1.4$& 0.5 \\
SO LAT (goal)& 145  &6.3 & $1.4$&0.4 \\
CMB-S4 & 150  &1.8 & $1.0$&0.7 \\
\hline
\end{tabular}
\caption{Noise specifications for the ground-based CMB experiments used in this analysis. Note that the exact noise specifications for CMB-S4 is yet to be finalised.}
\label{tab:mission_specs}
%\vspace{-0.5cm}
\end{table}

While forecasting the parameter constraints from upcoming CMB facilities, we use different combinations of $\tau$ and kSZ probes. For the $\tau$ measurements, it is expected that the low-$l$ $E$-mode polarization from the space-based mission LiteBIRD will be able to measure it at the cosmic variance limit where $\sigma_{\tau}^{\mathrm{obs}} = 0.002$ \citep{Suzuki2018}. For forecasting the kSZ signal, we compute the variance as
\bear
\left(\sigma^{\mathrm{kSZ}}_l\right)^2 &= \frac{2}{f_{\mathrm{sky}}(2l+1)} 
\left(D^{p}_l + D^{\mathrm{tSZ}}_l + D^{\mathrm{kSZ,tot}}_l \right.
\nline
& \left.+ D^{\mathrm{PS}}_l + D^{\mathrm{FG}}_l + N_l\right)^2, 
\ear
where the terms on the right hand side are: $D^{p}_l$ is the primary CMB (including lensing), computed for the best-fit cosmological parameters from \citet{PlanckCollaboration2016} using CAMB \citep{Lewis2000},
the thermal Sunyaev-Zeldovich (tSZ) component is taken as $D^{\mathrm{tSZ}}_{l=3000}= 4.4\,\mu \mathrm{K}^2$ \citep{George2015},
$D^{\mathrm{kSZ,tot}}_l$ is the total kSZ signal, with the OV part taken as $D^{\mathrm{kSZ, OV}}_{l=3000} = 2\, \mu \mathrm{K}^2$ \citep{Shaw2012},
the Poisson power spectrum is taken as $D^{\mathrm{PS}}_l= 7.59 \,\mu \mathrm{K}^2$ \citep{Reichardt2020}, 
the contamination from foregrounds is taken as $D^{\mathrm{FG}}_l\sim 12\, \mu \mathrm{K}^2$ \citep{Ade2019}, 
$N_l$ is the instrument noise which are specific for different missions given in Table \ref{tab:mission_specs} and
$f_{\mathrm{sky}}$ is the sky-fraction over which the signal is observed given in Table \ref{tab:mission_specs}.
We bin the power spectrum with $\Delta l=300$ at the central value $l=3000$ so as to decrease the variance on the measured signal. 

For different combination of the data sets, we obtain the posterior distribution of the parameters $\mathbf{\theta} \equiv \left\{\log(\zeta_0), \log(M_{\mathrm{min}, 0}), \alpha_{\zeta}, \alpha_M \right\}$ using publicly available Markov chain Monte Carlo (MCMC) sampler called \texttt{emcee} \citep{Foreman-Mackey2013}. The likelihood function is computed as ${\cal L}(\mathbf{\theta}) \propto \exp\left[-\chi^2(\mathbf{\theta}) / 2\right]$ and
\be
\chi^2(\mathbf{\theta}) = \left(\f{\tau(\mathbf{\theta}) - \tau^{\mathrm{obs}}}{\sigma_{\tau}^{\mathrm{obs}}}\right)^2
+ \left(\f{D_{l=3000}^{\mathrm{kSZ, tot}}(\mathbf{\theta}) - D^{\mathrm{kSZ, obs}}_{l=3000}}{\sigma^{\mathrm{kSZ, obs}}_{l=3000}}\right)^2.
\ee
All the free parameters are assumed to have flat priors with the range given in Table \ref{tab:current}. The priors on $M_{\mathrm{min}, 0}$ have been restricted to $10^7 - 10^{11} \Msun$, which covers the most interesting range of halo masses that can host star-forming galaxies. For example, the haloes where the gas can cool by atomic transitions have masses $M_{\mathrm{min}} \sim 10^8 \Msun$, while the effect of radiative feedback from reionization can increase $M_{\mathrm{min}}$ to $\sim 10^9 \Msun$ \citep[see, e.g.,][]{Choudhury2008}. Also note that we restrict $\alpha_M \leq 0$ which is because feedback processes will increase $M_{\mathrm{min}}$ with decreasing redshift. Further, we allow only those histories where reionization completes at $z > 5$, consistent with present constraints from Lyman-$\alpha$ optical depths \citep{McGreer2011,Kulkarni2019,Choudhury2020,Qin2020}.

\begin{figure}
  \includegraphics[width=0.45\textwidth]{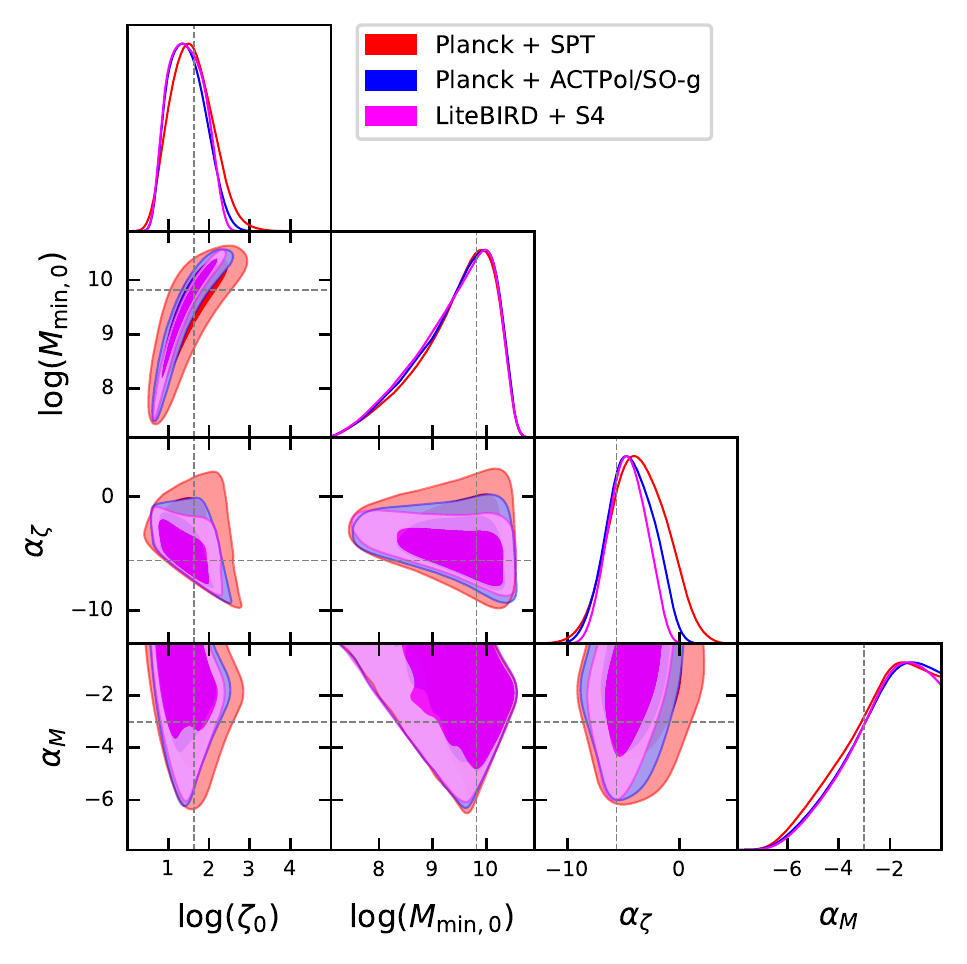}
  \caption{The marginalized posterior distribution of the model free parameters for different combinations of data sets as mentioned in the figure legend. We show the $68\%$ and $95\%$ contours in the two-dimensional plots. The corresponding constraints can be found in Table \ref{tab:current}. The dotted lines denote the input value used for the forecasting (the Planck+ACTPol/SO-g and LiteBIRD+S4 cases).} 
  \label{fig:3model_getdist_triangle}
%\vspace{-0.5cm}
\end{figure}

\begin{figure}
  \includegraphics[width=0.45\textwidth]{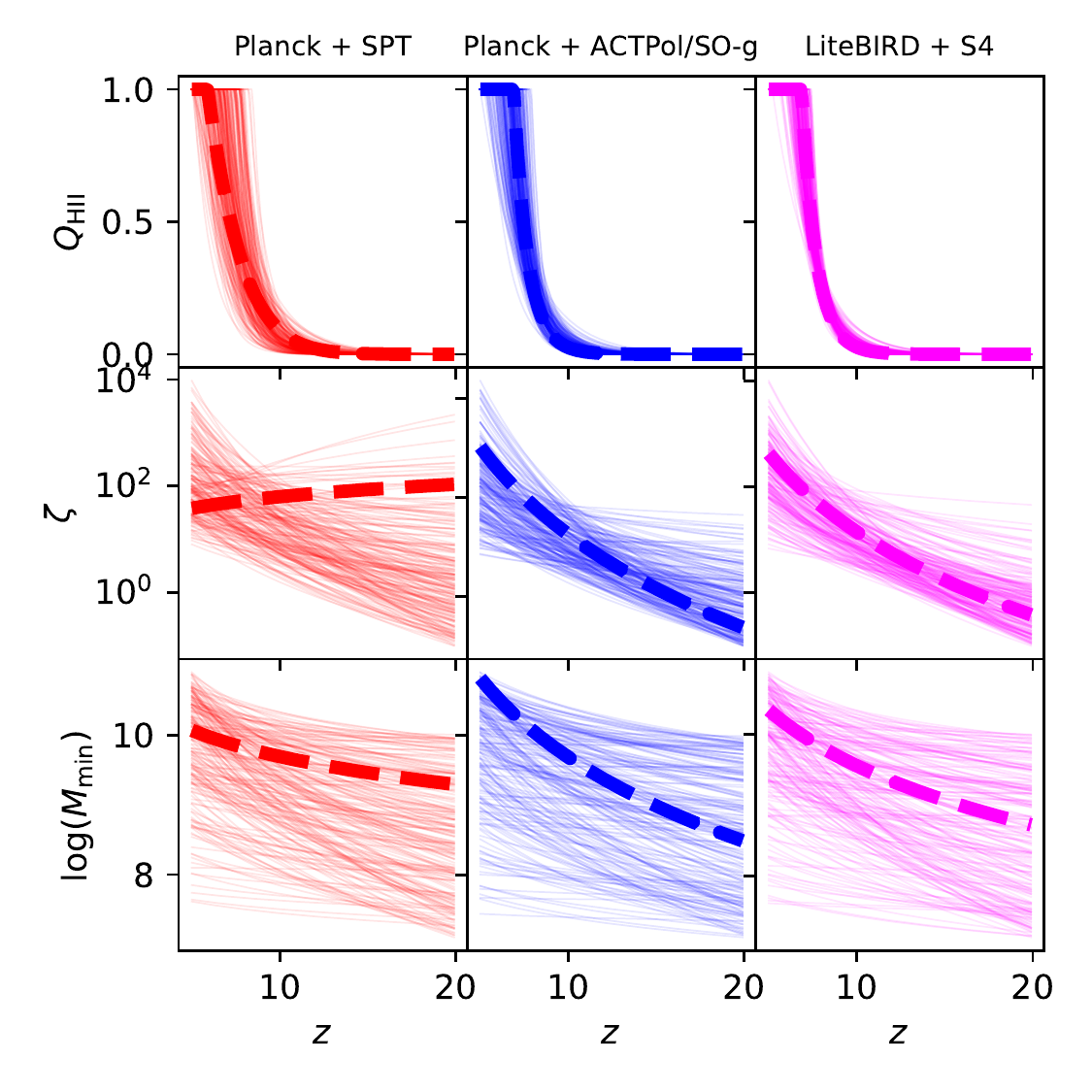}
  \caption{The redshift evolution of the ionized mass fraction $Q_{\mathrm{HII}}$ (top), the ionizing efficiency $\zeta$ (middle), and the minimum halo mass $M_{\mathrm{min}}$ that are capable to producing ionizing photons (bottom) for $200$ random samples from the MCMC chains. Different columns represent different combinations of data sets as mentioned in the figure. The thick dashed line corresponds to the best-fit model in each case.} 
  \label{fig:3model_getdist_z}
%\vspace{-0.5cm}
\end{figure}

%\vspace{-0.5cm}
\section{Results}

\subsection{Current constraints (Planck + SPT)}

\begin{table}
\begin{tabular}{|c|c|c|c|}
\hline
Data &  & Planck & Planck + SPT \\
Parameter & Prior & 68\% limits & 68\% limits\\
\hline
$\log(\zeta_0)$           & $[0, \infty]$ & $1.56^{+0.46}_{-0.58}$ & $1.58^{+0.44}_{-0.57}$ \\ %\\
$\log(M_{\mathrm{min},0})$ & $[7.0, 11.0]$ & $9.45^{+0.89}_{-0.36}$ & $9.44^{+0.88}_{-0.36}$ \\ %\\
$\alpha_{\zeta}$          & $[-\infty, \infty]$ & $-3.7 \pm 2.4$ & $-3.6 \pm 2.5$ \\ %\\
$\alpha_{M}$              & $[-\infty, 0]$ & $> -2.87$ & $> -2.95$ \\ %\\
%\hline
$\tau$                    & & $0.0558\pm 0.0066$ & $0.0563\pm 0.0064$ \\ %\\
$\Delta z$               & & $1.29^{+0.18}_{-0.58}$ & $1.30^{+0.19}_{-0.60}$ \\ %\\
$b^2_{\mathrm{kSZ}}  \times 10^7$    &  & $3.61^{+0.61}_{-0.47}$ & $3.61^{+0.63}_{-0.46}$ \\ %\\
$D_{200}^{BB} (\mathrm{n K}^2)$ & & $6.7^{+1.1}_{-3.5}$ & $6.8^{+1.1}_{-3.4}$ \\ %\\
\hline
\end{tabular}
\caption{Parameter constraints obtained from the MCMC-based analysis for the presently available data. The first four rows correspond to the free parameters of the model while the others are the derived parameters. The free parameters are assumed to have uniform priors in the range mentioned in the second column.}
\label{tab:current}
%\vspace{-0.3cm}
\end{table}

The parameter constraints obtained using Planck \citep{PlanckCollaboration2018} and SPT \citep{Reichardt2020} are shown in Table \ref{tab:current} with the one and two-dimensional posterior distributions shown in \fig{fig:3model_getdist_triangle} (the red contours and curves). For reference, we also show the constraints obtained using only Planck (i.e., ignoring the kSZ measurements from SPT) in  Table \ref{tab:current}. One can see that the constraints are very similar for the two cases, indicating that they are essentially driven by the small errors on $\tau$. 

From the table, we see that the data mildly prefers $M_{\mathrm{min}, 0} \gtrsim 10^9 \Msun$ (at $68\%$ C.L.). This is indicative of the fact that the radiative feedback processes are effective at $z \sim 8$ and hence the $M_{\mathrm{min}}$ is larger than that corresponding to simply atomically cooled haloes. The constraints also seem to favour $M_{\mathrm{min}, 0} < 10^{10.3} \Msun$ ($10^{10.6} \Msun$) at $68\%$ ($99\%$) C.L., thus ruling out reionization by extremely rare sources. Although the constraints on the individual free parameters are not stringent, from the contour plots in \fig{fig:3model_getdist_triangle}, we find that a substantial area in the $\log(\zeta_0) - \log(M_{\mathrm{min}, 0})$ plane is ruled out. The strong degeneracy between the two parameters does not allow stringent constraints on each of them. Although the parameter $\alpha_{\zeta}$ has large uncertainties, it slightly prefers negative values, thus indicating that the sources become more efficient in producing ionizing photons with time. This could be indicative of more efficient cooling and star formation and/or increased escape fraction. The above facts can also be confirmed from \fig{fig:3model_getdist_z} where we show the evolution of $Q_{\mathrm{HII}}$ (top), $\zeta$ (middle) and $\log(M_{\mathrm{min}})$ (bottom) for 200 randomly chosen models from the MCMC chains. The left hand panels correspond to the Planck+SPT case. It is clear that a wide range of values of $M_{\mathrm{min}}$ and $\zeta$ are allowed by the present data, however, the two parameters always combine in a way to provide reasonably tight constraints on $Q_{\mathrm{HII}}$.

\begin{figure}
  \includegraphics[width=0.45\textwidth]{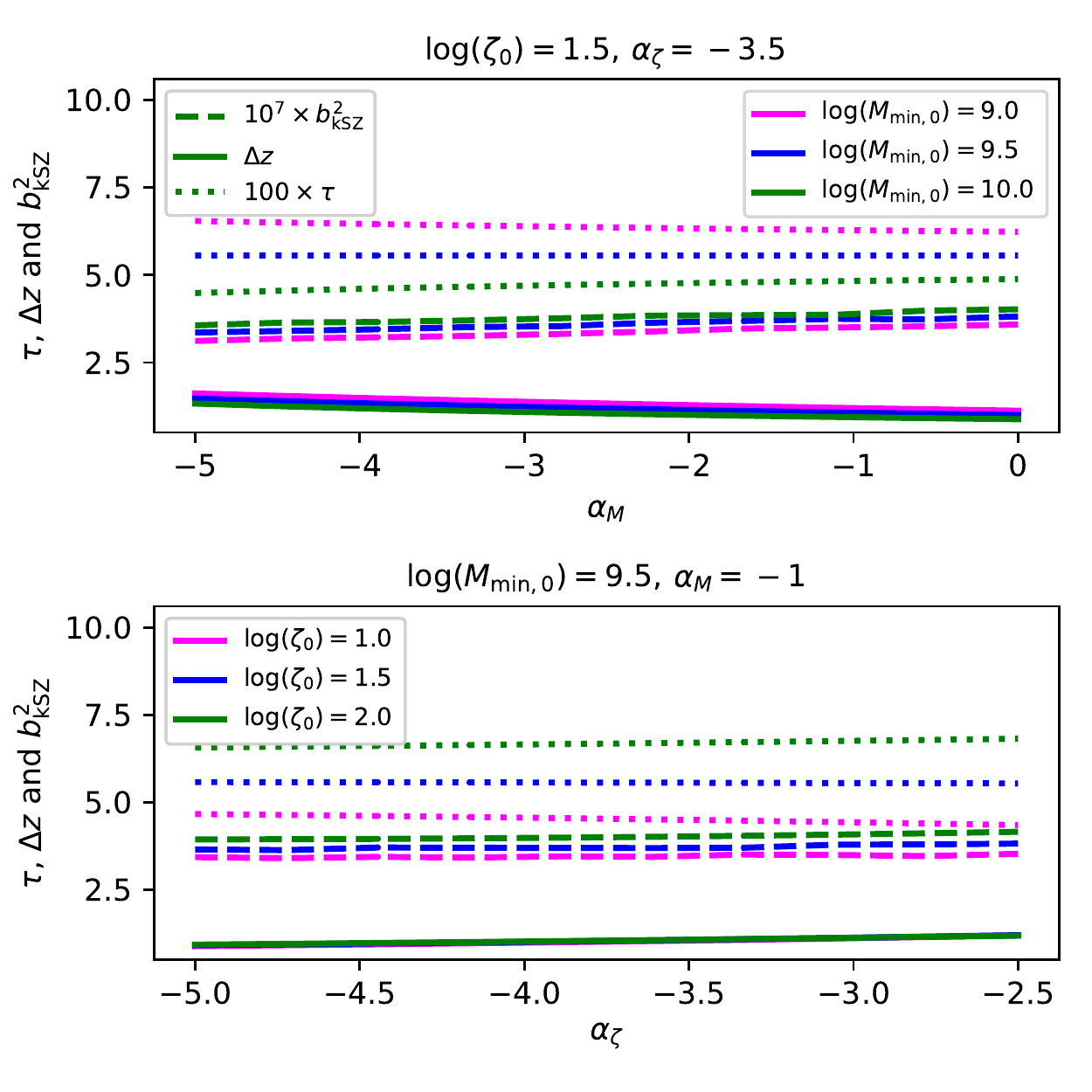}
  \caption{The dependence of the derived parameters $\tau$, $\Delta z$ and $b^2_{\mathrm{kSZ}}$ on the model parameters. In the top panel, we show the dependence on $\alpha_M$ for three values of $M_{\mathrm{min}, 0}$, while in the bottom panel we show the same on $\alpha_{\zeta}$ for three values of $\zeta_0$.} 
  \label{fig:plot_parameter_variation}
%\vspace{-0.5cm}
\end{figure}

In addition to the free parameters $\mathbf{\theta}$, we also show limits obtained on various derived quantities in Table \ref{tab:current}. Of particular interest are $\tau$, the reionization width $\Delta z \equiv z_{0.25} - z_{0.75}$ and the kSZ bias parameter \citep[introduced in our earlier work][]{Paul2020}
\be
b^2_{\mathrm{kSZ}}  \equiv \frac{1}{z_{0.01} - z_{0.99}} \int_{z_{0.99}}^{z_{0.01}} dz~\frac{P_{q_\perp}(k=l / \chi(z), z)}{P_{\rm DM}(k=l / \chi(z), z)},
\label{bkSZ}
\ee
where $z_{X}$ is the redshift where $Q_{\mathrm{HII}}= X$ and the integral is evaluated at $l = 3000$ (corresponding to the kSZ measurements). The dependencies of these derived parameters on the model parameters $\mathbf{\theta}$ are shown in \fig{fig:plot_parameter_variation}. Note that $b^2_{\mathrm{kSZ}}$, which measures the patchiness in the ionization field, is sensitive to both $M_{\mathrm{min}}$ (see the top panel of \fig{fig:plot_parameter_variation}) and $\zeta_0$ (bottom panel of the same figure). The one and two-dimensional posterior distributions of these three quantities are shown in \fig{fig:3model_getdist_derived_triangle} (the red contours).

From Table \ref{tab:current}, we find that the derived value of $\tau$ is slightly higher than that measured by Planck and the error is marginally smaller. This is due to the fact that our priors do not allow for reionization completing at $z < 5$, thus excluding scenarios with extremely small values of $\tau$. Our constraints on $\Delta z$ are more stringent than that of, e.g., \citet{Reichardt2020} and can put limits $\Delta z < 2.9$ at $99\%$ C.L. Larger values of $\Delta z$ would require either reionization completing at $z < 5$ or $\tau$-values larger than what is allowed by Planck. The constraints on $b^2_{\mathrm{kSZ}} = \left(3.61^{+0.61}_{-0.47}\right) \times 10^{-7}$ are indicative of the $M_{\mathrm{min}}$ range allowed by the data. 

We can also use the model to calculate the secondary $B$-mode polarization arising from patchy reionization due to scattering \citep[see, e.g.,][]{Dvorkin2009}. The present constraints on the $B$-mode polarization power spectrum from patchy reionization  $D^{BB}_{l=200} \equiv l(l+1)C^{BB}_{l=200}/2\pi$ are given in Table \ref{tab:current}. Interestingly, we find that $D^{BB}_{l=200} < 18$ nK$^2$ ($99\%$ C.L.). The presence of the $B$-mode signal from patchy reionization has consequences for the detection of the primordial gravitational waves \citep[for more details on this aspect, see][]{Mukherjee2019}. 

%\vspace{-0.5cm}
\subsection{Forecasts}

\begin{figure}
  \includegraphics[width=0.45\textwidth]{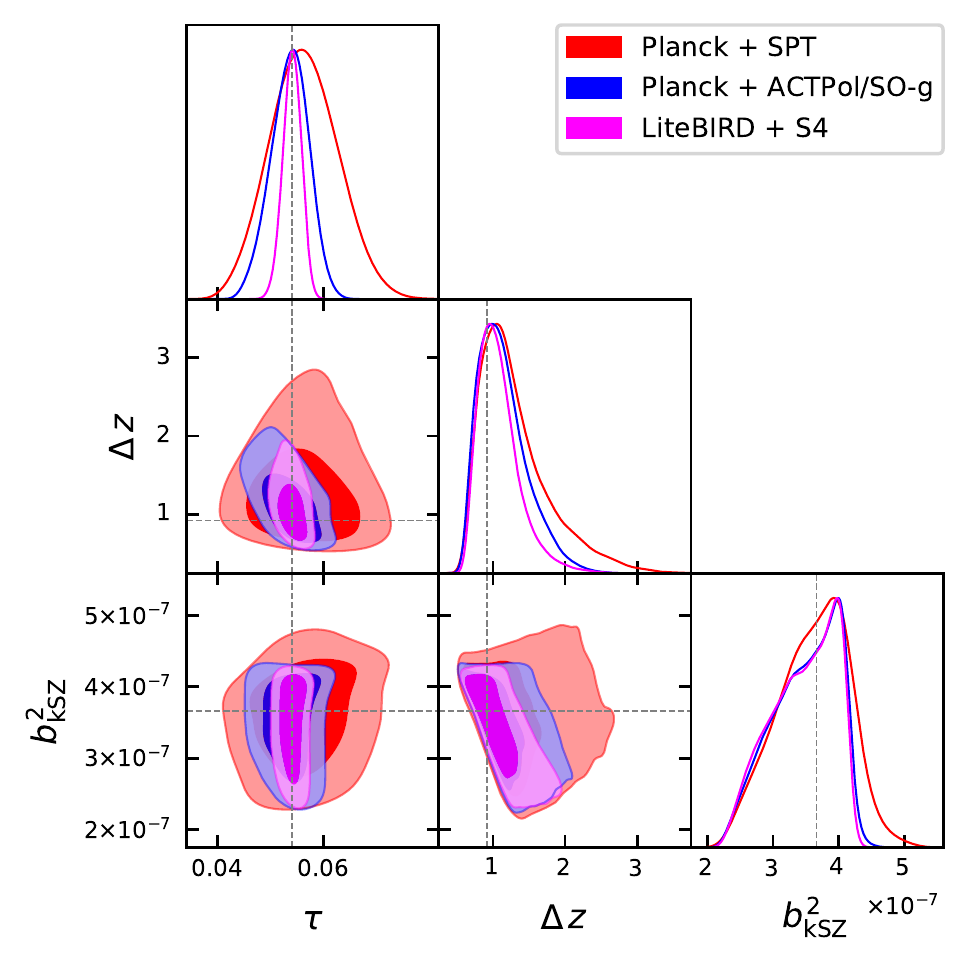}
  \caption{Same as \fig{fig:3model_getdist_triangle} but for the three derived parameters, namely, the optical depth $\tau$, the width of reionization $\Delta z$, and the kSZ bias parameter $b^2_{\mathrm{kSZ}}$.}
  \label{fig:3model_getdist_derived_triangle}
%\vspace{-0.5cm}
\end{figure}

We next study how the current constraints on reionization can be improved with upcoming CMB experiments. Combining the measurement of kSZ signal from the presently operating ground-based Adv-ACTPol \citep{Henderson2016} with Planck can already restrict the parameter space as can be seen from Table \ref{tab:forecasts}. Since the errors on the kSZ signal from the upcoming SO for $\sim 150$~GHz is similar to that of ACTPol, the results obtained from the two experiments are identical. Hence we denote the corresponding results as Planck+ACTPol/SO-g. As expected, the errors on all the parameters should decrease compared to the present constraints. The same can also be seen from the posterior distributions in \fig{fig:3model_getdist_triangle} and \ref{fig:3model_getdist_derived_triangle} (the blue contours and curves). Interestingly, introducing the ACTPol/SO-g in the analysis reduces the errors on $\tau$ to $\sim 0.004$, significantly smaller than the present errors from Planck. This represents the best constraints expected on $\tau$ before LiteBIRD is launched. Consequently, we can see from \fig{fig:3model_getdist_z} that the range of reionization histories would be significantly restricted. We also find that the constraints on $\zeta(z)$ to be more stringent (middle panel of \fig{fig:3model_getdist_triangle}) than the present ones.

\begin{table}
\begin{tabular}{|c|c|c|c|c|}
\hline
Data & & Planck & LiteBIRD \\
     & & + ACTPol/SO-g & + S4\\
Parameter & Input & 68\% limits & 68\% limits \\
\hline
$\log(\zeta_0)$           & $1.12$ & $1.46^{+0.37}_{-0.54}$ & $1.45^{+0.40}_{-0.49}$ \\ %\\
$\log(M_{\mathrm{min},0})$  & $8.94$ & $9.42^{+0.94}_{-0.39}$ & $9.42^{+0.92}_{-0.40}$ \\ %\\
$\alpha_{\zeta}$          & $-3.65$ & $-4.3\pm 2.0$ & $-4.6\pm 1.6$ \\ %\\
$\alpha_{M}$              & $-1.19$ & $> -2.76$ & $> -2.75$ \\ %\\
%\hline
$\tau$                   & $0.054$ & $0.0536^{+0.0038}_{-0.0032}$ & $0.0540\pm 0.0017$ \\ %\\
$\Delta z$               & $1.18$ & $1.14^{+0.19}_{-0.41}$ & $1.08^{+0.15}_{-0.33}$ \\ %\\
$b^2_{\mathrm{kSZ}}  \times 10^7$ & $3.66$ & $3.49^{+0.66}_{-0.33}$ & $3.46^{+0.67}_{-0.32}$ \\ %\\
$D_{200}^{BB} (\mathrm{n K}^2)$ & $4.01$ & $5.4^{+1.2}_{-2.0}$ & $5.2^{+1.0}_{-1.8}$ \\ %\\
\hline
\end{tabular}
\caption{Forecasts on various parameters for the upcoming facilities. The first four rows correspond to the free parameters of the model while the others are the derived parameters. The free parameters are assumed to have the same priors as mentioned in Table \ref{tab:current}. The second column shows the input values used to construct the default model based on which the forecasts are made.}
\label{tab:forecasts}
%\vspace{-0.3cm}
\end{table}

The constraints would be must more stringent when we combine the kSZ measurements from the upcoming ground-based CMB experiments such as SO and CMB-S4 along with $\tau$ measurement from LiteBIRD. The results are shown in magenta in \fig{fig:3model_getdist_triangle} and \ref{fig:3model_getdist_derived_triangle}. Unsurprisingly, the uncertainties on $\tau$ approach the cosmic variance limits. The standard deviation on $\Delta z$ is $\sim 0.3$, almost half the present value (which is $0.5$). The allowed ranges of the parameters $\zeta(z)$ and $M_{\mathrm{min}}(z)$ are also considerably reduced. For example, if we assume that our chosen input model indeed represents the true model (which need not necessarily be the case), we can rule out $\alpha_{\zeta} > 0$ at $> 99\%$ C.L., thus implying that the reionization sources become more efficient with time. Interestingly, we find the standard deviation on $D_{l=200}^{BB}$ to be $1.43$nK$^2$, significantly smaller than the present bounds. Again, assuming our input model represents the true case, the upper limit on $D_{l=200}^{BB}$ is $8$nK$^2$ ($99\%$ C.L.). This should lead to $\lesssim 10\%$ bias in the value of $r= 10^{-3}$ \citep{Mukherjee2019}.

%\vspace{-0.6cm}
\section{Discussions}
\label{sec:conc}
%\vspace{-0.1cm}

Using a physical semi-numerical model of reionization (\texttt{SCRIPT}), we constrain the reionization history by comparing the predictions with \emph{only CMB observables}. In particular, we use the measurement of optical depth from Planck \citep{PlanckCollaboration2018} and the kSZ measurement from SPT \citep{Reichardt2020} to obtain the constraints. Our model has for free parameters ($\zeta_0, \alpha_{\zeta}, M_{\mathrm{min}, 0}, \alpha_M$) which characterize the redshift evolution of the ionization efficiency $\zeta(z)$, and the minimum halo mass $M_{\mathrm{min}}(z)$ that can produce ionizing photons. The main results of the analysis are:

%\vspace{-0.2cm}
\begin{itemize}
\item We constrain the duration of reionization $\Delta z=1.30^{+0.19}_{-0.60}$ and limit $\Delta z < 2.9$ at $99\%$ C.L. Our limits are consistent with but more stringent than the measurement from SPT \citep{Reichardt2020}.
\item Our analysis mildly favours $M_{\mathrm{min}} \gtrsim 10^9 \Msun$ ($68\%$ C.L.) at $z \sim 8$, thus indicating presence of radiative feedback at these redshifts.
\item The kSZ bias parameter is constrained to $b^2_{\rm kSZ}= \left(3.61^{+0.61}_{-0.47}\right) \times 10^{-7}$, which indicates that the patchiness in the electron density during the epoch of reionization cannot be extremely large. This also implies that reionization cannot be driven by extremely rare sources.
\item Another important implication of these results is it provides the first upper bound from observations on the $B$-mode polarization signal which can be produced due to patchy reionization: $D_{l=200}^{BB}< 18$ nK$^2$ at $99\%$ C.L. This has important implications for the detection of the primordial gravitational waves.
\end{itemize}

%\vspace{-0.2cm}
In addition to the present constraints, we have also studied the possible improvements in the parameter limits with upcoming CMB experiments. Our analysis allows for constraints using a generalized parametrization of reionization and can be useful in predicting the signal expected with the future experiments, e.g., $B$-mode polarization and the redshifted 21~cm observations. In future, we plan to extend our analysis taking into account all the other available data sets related to reionization and obtain bounds on the allowed reionization models.

%\vspace{-0.7cm}
\section*{Acknowledgements}
%\vspace{-0.2cm}
SM acknowledges useful discussions with Joseph Silk and Benjamin D. Wandelt. TRC acknowledges support of the Department of Atomic Energy, Government of India, under project no. 12-R\&D-TFR-5.02-0700. SM is supported by the research program Innovational Research Incentives Scheme (Vernieuwingsimpuls), which is financed by the Netherlands Organization for Scientific Research through the NWO VIDI Grant No. 639.042.612-Nissanke. SP acknowledges SARAO for support through the SKA postdoctoral fellowship.

%\vspace{-0.7cm}
\section*{Data availability}
%\vspace{-0.2cm}
The data underlying this article will be shared on reasonable request to the corresponding author (TRC).

%\vspace{-0.5cm}
\bibliographystyle{mnras}
\bibliography{CMB_reion}

\end{document}